\documentclass[iop]{emulateapj}

\usepackage{lineno, amsmath, appendix}

\newcommand\kms{\ifmmode{\rm km\thinspace s^{-1}}\else km\thinspace
 s$^{-1}$\fi} 
\newcommand\hstar{HD~174881}

\shortauthors{Torres et al.}
\shorttitle{\hstar}

\begin{document}
\submitted{Accepted for publication in The Astrophysical Journal}

\title{Absolute Dimensions of the Interferometric Binary \hstar: A Test
  of Stellar Evolution Models for Evolved Stars}

\author{
Guillermo Torres\altaffilmark{1}, 
Andrew F.\ Boden\altaffilmark{2}, 
John D.\ Monnier\altaffilmark{3}, and 
Gerard T.\ van Belle\altaffilmark{4} 
}

\altaffiltext{1}{Center for Astrophysics $\vert$ Harvard \&
  Smithsonian, 60 Garden St., Cambridge, MA 02138, USA; gtorres@cfa.harvard.edu}

\altaffiltext{2}{California Institute of Technology, Mail Code 11-17,
  1200 East California Boulevard, Pasadena, CA 91125, USA}

\altaffiltext{3}{
Astronomy Department, University of Michigan, Ann Arbor, MI 48109, USA}

\altaffiltext{4}{Lowell Observatory, 1400 West Mars Hill Road, Flagstaff, AZ 86001, USA}

\begin{abstract} 
We report high-resolution spectroscopic monitoring and
long-baseline interferometric observations with the PTI of the
215-day binary system \hstar\ (\ion{K1}{2--iii}),
composed of two giant stars. The system is spatially resolved
with the PTI, as well as in archival measurements with the
CHARA Array. Our analysis of these observations, along with
an analysis of the spectral energy distribution, have allowed
us to infer accurate values for the absolute masses
($3.367^{+0.045}_{-0.041}~M_{\sun}$ and $3.476^{+0.043}_{-0.043}~M_{\sun}$),
radii ($34.0 \pm 1.3~R_{\sun}$ and $22.7 \pm 1.8~R_{\sun}$),
effective temperatures ($4620 \pm 100$~K and $4880 \pm 150$~K),
and bolometric luminosities of both components,
as well as other properties including the orbital parallax (distance).
These provide valuable tests of stellar evolution
models for evolved stars, which are still relatively uncommon
compared to the situation for main-sequence stars.
We find generally good agreement of all of these properties
of \hstar\ with two sets of recent models (MIST, and PARSEC)
at compositions near solar, for ages of 255--273~Myr.
We also find evidence of an infrared excess, based largely on
the flux measurements from IRAS at 60 and 100\thinspace$\mu$m.
\end{abstract}



\section{Introduction}
\label{sec:introduction}

Stellar evolution theory has had remarkable success in reproducing the
observed properties of stars over much of the hydrogen-burning main
sequence \citep[see, e.g.,][]{Andersen:1991a, Torres:2010}. This has
been made possible by significant advances in our knowledge of stellar physics,
aided by the ever growing observational constraints on models provided 
binary systems. Those objects allow direct and precise determinations
of their component masses (the most fundamental stellar property), as well as 
their radii, temperatures, chemical compositions, and other important properties.
Traditionally,
the best measurements have been made in double-lined eclipsing
systems, although many spectroscopic-astrometric binaries have
provided valuable information as well.

Models for post-main sequence stars, on the other hand, are somewhat
less secure, due in part to the fewer empirical constraints currently
available. Binaries with well-detached giant or
subgiant components, as needed for meaningful tests of theory, are
much less common compared to similar main-sequence systems. To
accommodate the larger stars, the orbits necessarily have longer
periods, which makes eclipses less likely. In favorable cases the
stars can be spatially resolved by astrometric techniques, such as
speckle interferometry or long-baseline interferometry. When
complemented by spectroscopy, if needed, this also provides a way to
determine the component masses, along with the orbital parallax.

Examples of binaries with post-main-sequence components that have provided valuable
constraints on stellar evolution theory include, among others, the
eclipsing systems TZ~For \citep{Andersen:1991b, Gallene:2016} and
HD~187669 \citep{Helminiak:2015}, the astrometric-spectroscopic binary
Capella \cite[$\alpha$~Aur;][]{Torres:2009, Torres:2015, Weber:2011},
and nearly two dozen other eclipsing systems in the Milky way and
in the Magellanic Clouds
\citep[e.g.,][]{Graczyk:2014, Graczyk:2018, Graczyk:2020, Rowan:2024}.

In this paper we report an analysis of the astrometric-spectroscopic
binary \hstar\ (HR~7112; \ion{K1}{2--iii}, $
V = 6.18$), a well-detached system
in which both components are giants. It was discovered
spectroscopically by \cite{Appleton:1995}, who presented a
double-lined orbit with a period of 215 days and a small eccentricity 
of $e = 0.141$.
For this study we have obtained additional, higher-resolution
spectroscopic observations at the Center for Astrophysics (CfA), which
have allowed us to improve the orbit significantly. We also pursued
the object with the Palomar Testbed Interferometer (PTI), and have
successfully spatially resolved the binary for the first time. The combination
of these measurements has allowed us to infer many properties of the
system that we use below, to provide stringent constraints on models
for the giant phase.

The layout of our paper is as follows. In
Section~\ref{sec:spectroscopy} we report our new spectroscopic
observations of \hstar, and report also other radial-velocity
measurements from the literature that we incorporate into our
analysis. Section~\ref{sec:interferometry} describes our
interferometric observations with the PTI, as well as additional
archival interferometric observations from the Center for High Angular
Resolution Astronomy (CHARA) Array that also resolve the binary.
Section~\ref{sec:sed} then presents our analysis of the spectral
energy distribution of the system, from which we infer the individual
absolute luminosities of the components and other properties.
In Section~\ref{sec:orbit} we
combine the spectroscopic and astrometric measurements to derive the
3D orbit. A comparison of the system properties against models of
stellar evolution is given in Section~\ref{sec:discussion}.
Our conclusions are drawn in Section~\ref{sec:conclusions}.

\section{Spectroscopic Observations and Reductions}
\label{sec:spectroscopy}

Our observations of \hstar\ at the CfA were conducted with an echelle
spectrograph on the 1.5m Wyeth reflector at the (now closed) Oak Ridge
Observatory (Massachusetts, USA), and occasionally also with a nearly
identical instrument on the 1.5m Tillinghast reflector at the
F.\ L.\ Whipple Observatory (Arizona, USA). A single echelle order
spanning 45\,\AA\ was recorded with intensified photon-counting
Reticon detectors at a central wavelength of 5187\,\AA, which includes
the \ion{Mg}{1}~b triplet.  The resolving power of these instruments
is $\lambda/\Delta\lambda \approx 35,\!000$.  A total of 81 spectra
were obtained from February 2000 to July 2004, with signal-to-noise
ratios of 47--75 per resolution element of 8.5~\kms.

Radial velocities for the two components were derived with the
two-dimensional cross-correlation algorithm TODCOR
\citep{Zucker:1994}, which allows velocities to be obtained reliably
even when the spectral lines are blended. This technique uses two
templates, one for each component of the binary. They were selected
from an extensive library of calculated spectra based on model
atmospheres by R.\ L.\ Kurucz \citep[see][]{Nordstrom:1994,
  Latham:2002}, and a line list manually tuned to better match real
stars.  These calculated spectra are available for a wide range of
effective temperatures ($T_{\rm eff}$), projected rotational
velocities ($v \sin i$), surface gravities ($\log g$), and
metallicities. Experience has shown that the radial velocities are
most sensitive to the rotational velocity and temperature adopted, and
less dependent on surface gravity and metallicity.  Consequently, we
first determined the optimum template for each star from grids of
cross-correlations over broad ranges in $T_{\rm eff}$ and $v \sin i$,
seeking to maximize the average correlation value weighted by the strength
of each exposure. The metallicity was held at the solar value, and
surface gravities were set to values appropriate for giant
stars. Subsequently, we explored the possibility of determining the
$\log g$ values as well, even though this has usually been very
difficult to do for double-lined spectroscopic binaries.  We repeated
the temperature and rotational velocity determinations at fixed values
of $\log g$ from 0.5 to 4.5 for each star, and found that there was a
distinct preference for surface gravities near 2.0 for both.
Interpolation yielded the final values of $\log g = 1.9 \pm 0.2$ for
the brighter of the two stars (hereafter star~A), which turns out
to be the \emph{less} massive one in the system, and $\log g = 2.0 \pm 0.3$
for the other (star~B).
The effective temperatures we determined for the
two components are $4620 \pm 100$~K and $4880 \pm 150$~K, respectively. The
rotational velocities that give the highest average correlation are $7
\pm 2$~\kms\ for star~A and $8 \pm 3$~\kms\ for star~B.
However, we caution that these values may well be overestimated, as
they are based on a comparison with synthetic spectra computed for a
macroturbulent velocity $\zeta_{\rm RT} = 1$~\kms\ (the only value
available in our template library) that is more
appropriate for dwarfs than giants. The $v \sin i$ values we have
derived may simply be compensating to some extent for the increased
line broadening from macroturbulence that is more common in more
luminous stars such as these. Indeed, \cite{DeMedeiros:1999} have
reported a rotational broadening of approximately 1~\kms\ for both
components of \hstar.
To measure the radial velocities, we adopted templates from our
library with parameters nearest to those reported above.  The
stability of the zero-point of our velocity system was monitored by
means of exposures of the dusk and dawn sky, and small run-to-run
corrections were applied in the manner described by
\cite{Latham:1992}.

In addition to the radial velocities, we derived the spectroscopic light ratio
between the two stars following \cite{Zucker:1994}. We obtained
$(F_{\rm B}/F_{\rm A})_{\rm sp} = 0.752 \pm 0.011$, corresponding to a
brightness difference of $\Delta m = 0.31 \pm 0.02$ mag at the mean
wavelength of our observations (5187\,\AA). Star~A is therefore
brighter and cooler, but less massive than star~B.

Due to the narrow wavelength coverage of the CfA spectra, there is the
potential for systematic errors in the velocities resulting from lines
of the stars moving in and out of the spectral window with orbital
phase \citep{Latham:1996}.  These errors are occasionally significant,
and experience has shown that this must be checked on a case-by-case
basis \citep[see, e.g.,][]{Torres:1997, Torres:2000}. For this, we
performed numerical simulations in which we generated artificial
composite spectra by adding together synthetic spectra for the two
components, with Doppler shifts appropriate for each actual time of
observation, computed from a preliminary orbital solution.  The light
ratio adopted is that reported above.  We then processed these
simulated spectra with TODCOR in the same manner as the real spectra,
and compared the input and output velocities. The differences were all
well below 1~\kms.  Nevertheless, we applied these differences as
corrections to the raw velocities, and the final velocities including
these adjustments are given in Table~\ref{tab:rvs}.  Similar
corrections were derived for the light ratio, and are already
accounted for in the value listed above.

\setlength{\tabcolsep}{2pt}
\begin{deluxetable}{lc@{~~~}ccc}
\tablewidth{0pc}
\tablecaption{CfA Radial Velocity Measurements for \hstar \label{tab:rvs}}
\tablehead{
\colhead{HJD} &
\colhead{Year~~~} &
\colhead{Phase} &
\colhead{$RV_{\rm A}$} &
\colhead{$RV_{\rm B}$}
\\
\colhead{(2,400,000+)} &
\colhead{} &
\colhead{} &
\colhead{(\kms)} &
\colhead{(\kms)}
}
\startdata
 51596.9077  &  2000.1421  &  0.8531  &   $-1.10 \pm 0.48$      &  $-33.98 \pm 0.55$\phn  \\
 51611.8966  &  2000.1832  &  0.9227  &  $-10.87 \pm 0.47$\phn  &  $-27.10 \pm 0.53$\phn  \\
 51627.8306  &  2000.2268  &  0.9968  &  $-23.21 \pm 0.45$\phn  &  $-14.96 \pm 0.51$\phn  \\
 51665.8192  &  2000.3308  &  0.1734  &  $-40.98 \pm 0.47$\phn  &    \phs$2.67 \pm 0.53$  \\
 51690.7850  &  2000.3991  &  0.2895  &  $-35.74 \pm 0.41$\phn  &   $-1.92 \pm 0.46$      
\enddata

\tablecomments{Orbital phases were calculated using the ephemeris in
  Table~\ref{tab:mcmc}. Star~A is the less massive star.
  (This table is available in its entirety in
  machine-readable form).}

\end{deluxetable}
\setlength{\tabcolsep}{6pt}

In addition to our own velocities of \hstar, a data set of similar
quality was reported by \cite{DeMedeiros:1999}, obtained with the
CORAVEL spectrometer on the 1m Swiss telescope at the Haute-Provence
Observatory (France). Separate spectroscopic orbital solutions with
our data and those of De~Medeiros give consistent velocity
semiamplitudes. We have therefore incorporated the CORAVEL data into
our analysis below.

\section{Interferometric Observations}
\label{sec:interferometry}

\subsection{PTI}

Near-infrared, long-baseline interferometric measurements of
\objectname[HD 174881]{HD~174881} and
calibration sources were conducted with the Palomar Testbed Interferometer
(PTI; \citep{Colavita:1999}) in the $H$ and $K$ bands ($\lambda\sim 1.6~\mu$m
and $\sim$ 2.2~$\mu$m, respectively) between 2000 and 2006.  The maximum
PTI baseline (110~m) provided a minimum $K$-band fringe spacing of approximately
4~mas, making the \hstar\ system readily resolvable.  

The PTI interferometric observable used for these measurements is the
fringe contrast or ``visibility'' (specifically, the power-normalized
visibility modulus squared, or $V^2$) of the observed brightness
distribution on the sky. \hstar\
was typically observed in conjunction with calibration
objects, and each observation (or scan) was approximately 130 seconds long.
As in previous publications, PTI $V^2$ data reduction and calibration
follow standard procedures described by \citet{Colavita:2003} and
\cite{Boden:1998}, respectively.  Observations of \hstar\ and associated
calibration sources
(\objectname[HD 173667]{HD~173667} and \objectname[HD 182488]{HD~182488})
resulted in 466 calibrated $K$-band visibility scans on a total of 76
nights spanning a period of nearly 6 years, or about 10 orbital periods.

\setlength{\tabcolsep}{8pt}
\begin{deluxetable*}{lcccccc}
\tablewidth{0pc}
\tablecaption{PTI Visibility Measurements for \hstar \label{tab:v2}}
\tablehead{
\colhead{JD$-$2,400,000} &
\colhead{Year} &
\colhead{Phase} &
\colhead{$\lambda$} &
\colhead{$V^2$} &
\colhead{$u$} &
\colhead{$v$}
\\
\colhead{} &
\colhead{} &
\colhead{} &
\colhead{($\mu$m)} &
\colhead{} &
\colhead{(m)} &
\colhead{(m)}
}
\startdata
  51667.9231  &  2000.3365  &   0.7637  &  2.2225  &  $0.1191 \pm 0.0100$  & $ -59.36425$  &  $-92.01252$ \\
  51667.9254  &  2000.3365  &   0.7641  &  2.2215  &  $0.1209 \pm 0.0100$  & $ -58.81588$  &  $-92.42476$ \\
  51667.9466  &  2000.3366  &   0.7680  &  2.2170  &  $0.1356 \pm 0.0116$  & $ -53.18509$  &  $-96.03746$ \\

  51667.9666  &  2000.3367  &   0.7617  &  2.2246  &  $0.2007 \pm 0.0231$  & $ -47.02468$  &  $-99.07473$ \\
  51677.9241  &  2000.3639  &   0.5891  &  2.2342  &  $0.4462 \pm 0.0219$  & $ -51.78284$  &  $-96.79841$ 
\enddata

\tablecomments{Orbital phases were calculated using the ephemeris in
  Table~\ref{tab:mcmc}. The $\lambda$ values correspond to the
  effective (flux-weighted) center-band wavelengths of the PTI
  passband. (This table is available in its entirety in
  machine-readable form).}

\end{deluxetable*}
\setlength{\tabcolsep}{6pt}

\subsection{CHARA}

The CHARA Array is the world's longest baseline optical/infrared interferometer, with six 1m telescopes spread across Mt.\ Wilson, California \citep{theo2005}.  The maximum baseline of 330m affords an angular resolution of $\Theta\sim \lambda/(2B_{\rm max})= 0.5$~mas, when observing in the near-infrared $H$ band ($\lambda\sim$1.65 $\mu$m).  

\hstar\ was observed in the $H$ band on the nights of UT\thinspace 2007\thinspace Jul\thinspace 04 and UT\thinspace 2007\thinspace Jul\thinspace 07, with the (then) recently-commissioned Michigan InfraRed Combiner \citep[MIRC;][]{mirc2004}.  At that time, MIRC could combine light from any four CHARA telescopes, measuring six baselines and four closure phases simultaneously.   These archival data were processed with an IDL pipeline that used conventional Fourier Transform techniques to extract the visibility and phases from the fringes created in the image plane.  The calibrated data were saved in the OI-FITS format \citep{oifits2005}, and will be deposited with the OI Database hosted at the Jean-Marie Mariotti Center (\url{https://jmmc.fr}).  

The first night of data (UT\thinspace 2007\thinspace Jul\thinspace 04) was rather limited, including only three CHARA telescopes (S1-W1-W2) and using the calibrator $\gamma$~Lyr (uniform-disk angular diameter $\phi_{\rm UD} = 0.737 \pm 0.15$~mas). The second night (UT\thinspace 2007\thinspace Jul\thinspace 07) had more and better quality data using four telescopes (S1-E1-W1-W2), with $\sigma$~Cyg as the primary calibrator ($\phi_{\rm UD} = 0.54 \pm 0.02$~mas).  Calibrator diameter estimates were based on a combination of an internal MIRC calibrator study (unpublished) and visible-light measurements by the PAVO instrument \citep{Maestro2013}.   

Fitting simultaneously for binary separation, flux ratio, and component diameters allowed us to measure the sizes of the individual components of \hstar.  Using only the data from the higher-quality UT\thinspace 2007\thinspace Jul\thinspace 07 dataset, we measured uniform-disk diameters of $0.817 \pm 0.030$~mas for star~A, $0.50 \pm 0.05$~mas for star~B, and a flux ratio (B/A) of $0.458 \pm 0.003$ in the $H$ band.  Table~\ref{tab:chara} contains the measured separations and position angles between the components from both dates. Note that the component sizes from UT\thinspace 2007\thinspace Jul\thinspace 07 were used as fixed values for fitting the binary model for UT\thinspace 2007\thinspace Jul\thinspace 04, as the earlier and smaller dataset could not constrain the sizes on its own.

\begin{figure}
\epsscale{1.4}
\includegraphics[width=4.5in, trim = 1.2in 0.8in 0.2in 0.2in, clip]{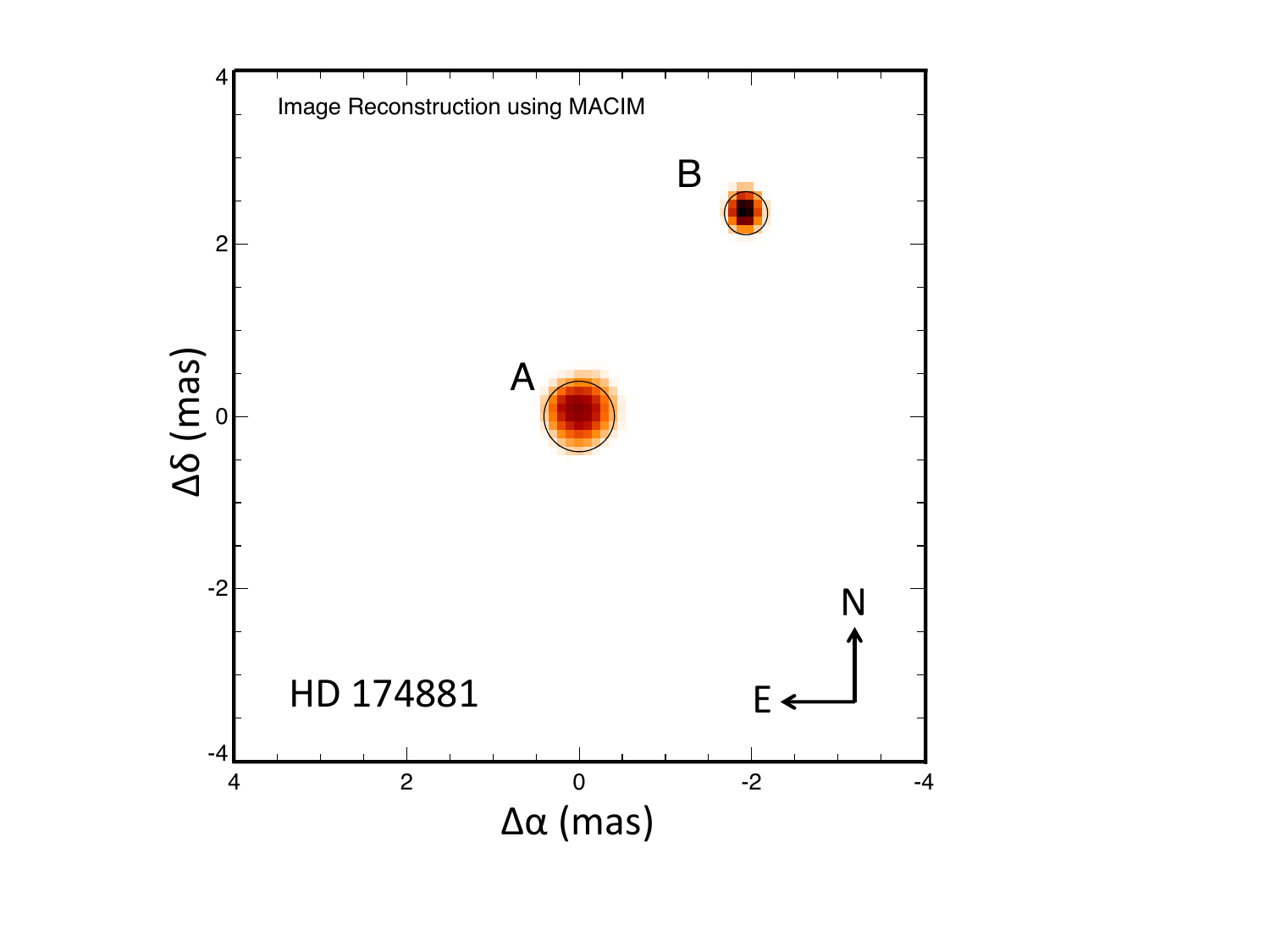}
\figcaption{Image reconstruction of \hstar\ from the MIRC data of UT\thinspace 2007\thinspace Jul\thinspace 07, using the MACIM algorithm. The location and uniform diameters from the model-fitting procedure are overlaid.\label{fig:image}}
\end{figure}

While model fitting is the most reliable and precise way to characterize the MIRC data, we also produced an image reconstruction of \hstar\ using the MACIM algorithm  \citep{macim2006}. Because of the highly sparse $uv$ coverage, we used an image prior of two Gaussians (with full width at half maximum of 0.5~mas) centered on the expected locations of the stars based on modeling. We further employed a “uniform disk” regularizer, which is the $\ell_\frac{1}{2}$-norm of the spatial gradient of the image, first described by \citet{baron2014}.  Figure~\ref{fig:image} shows the MACIM image along with diameter circles from model fitting. The agreement between model and image is excellent. A more detailed comparison between the observations and the
model is provided in the Appendix.

\setlength{\tabcolsep}{4pt}
\begin{deluxetable}{lcccccc}
\tablewidth{0pc}
\tablecaption{CHARA Measurements for \hstar \label{tab:chara}}
\tablehead{
\colhead{MJD} &
\colhead{UT Date} &
\colhead{$\theta$} &
\colhead{$\rho$} &
\colhead{$\sigma_{\rm maj}$} &
\colhead{$\sigma_{\rm min}$} &
\colhead{$\psi$}
\\
\colhead{} &
\colhead{} &
\colhead{(deg)} &
\colhead{(mas)} &
\colhead{(mas)} &
\colhead{(mas)} &
\colhead{(deg)}
}
\startdata
54285.326  &  2007 Jul 04  &  315.1\phn  &  3.09\phn  &  0.216  &  0.04\phn\phn  &  45.1\phn   \\
54288.250  &  2007 Jul 07  &  320.64     &  3.051     &  0.015  &  0.0149        &  50.64      
\enddata

\tablecomments{Columns $\sigma_{\rm maj}$ and $\sigma_{\rm min}$ represent the
major and minor axes of the 1-$\sigma$ error ellipse for each
measurement, and $\psi$ gives the orientation of the major axis
relative to the direction to the north.  Position angles are referred
to the International Celestial Reference Frame (effectively J2000).}
\end{deluxetable}
\setlength{\tabcolsep}{6pt}

\section{Spectral Energy Distribution}
\label{sec:sed}

A spectral energy distribution (SED) analysis of the \hstar\
system was used to infer the radiometric parameters (e.g., effective temperature,
bolometric flux $f_{\rm bol}$, and eventually luminosity, when combined with system distance)
for each of the components.  Flux inputs to the SED modeling
presented here are the large collection of archival combined-light photometry and
flux measurements available from the literature, in the following
photometric systems: Johnson \citep{Mermilliod:1987}, DDO \citep{McClure:1981},
Straizys \citep{Straizys:1989}, 2MASS \citep{Cutri:2003}\footnote{An
additional $K_{\rm S}$-band measurement was obtained for this work, and is
described below in Section~\ref{sec:discussion}.},
AKARI \citep{Murakami:2007}, WISE \citep{Cutri:2012}, and IRAS \citep{Beichman:1988}.
Additionally, we used the Gaia BP/RP low-dispersion spectra \citep{GaiaDR3:2023},
with corrections as recommended by \cite{Huang:2024}, and importantly also,
the in-band component flux
ratios. The latter were derived from the system optical spectra at 5187~\AA,
and the interferometric observables in the $H$ and $K$ bands from CHARA
and PTI, respectively, as described in the preceding sections.
These flux and flux ratio data were jointly analyzed with a custom two-component SED
modeling code introduced in the work of \cite{Boden:2005}, using 
solar metallicity PHOENIX model spectra from \cite{Husser:2013}.
The model atmospheres underlying these spectra
adopt spherical geometry for the stellar structure.

\begin{figure}
\epsscale{1.18}
\plotone{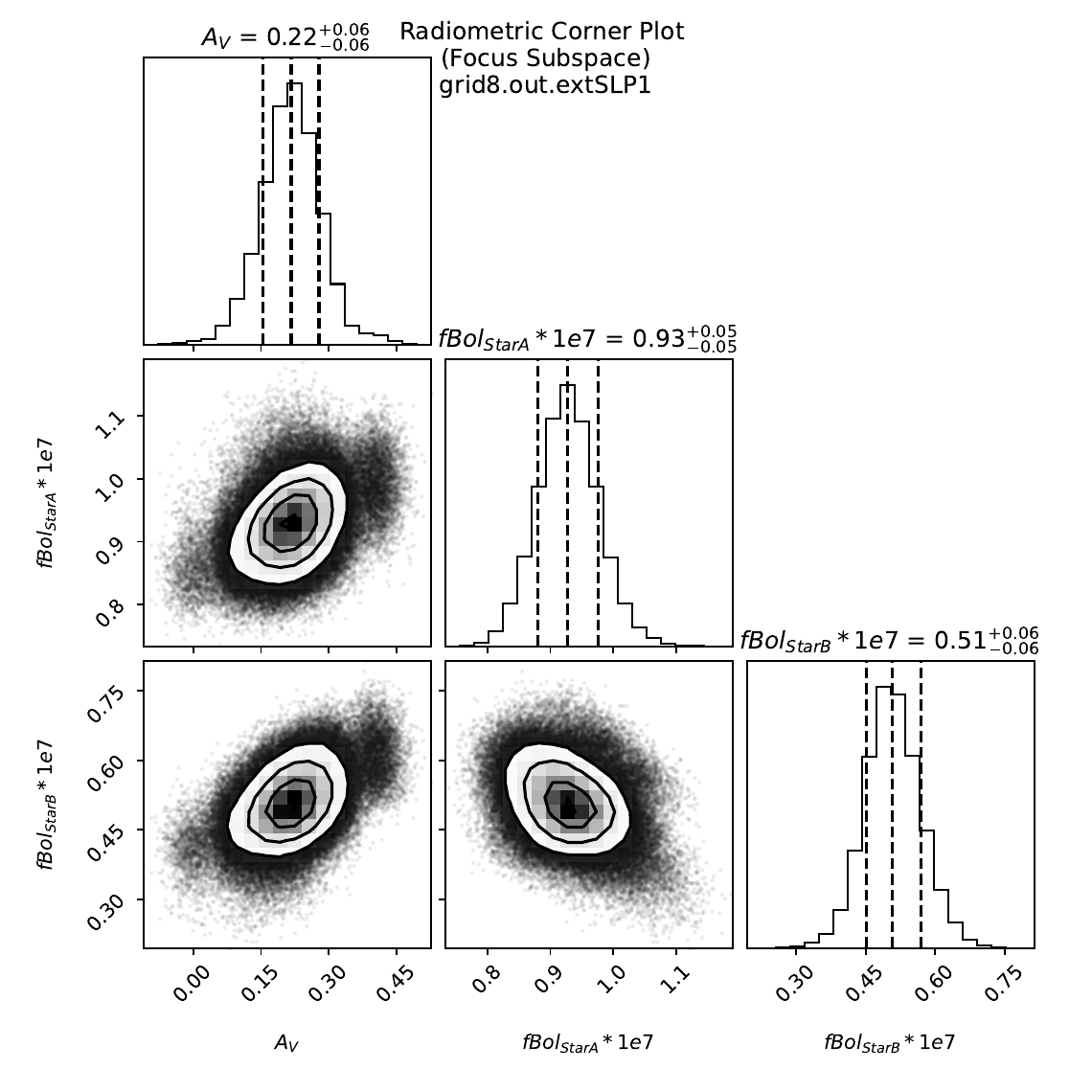}
\figcaption{Corner plot showing the correlations among a selection
of the fitted parameters of the radiometric analysis.\label{fig:sedcorner}}
\end{figure}

Numerical quadratures of the resulting component SED models directly
yield estimates for the bolometric fluxes of the individual components.
Our modeling
suggested that modest extinction along the line of sight is
necessary to reproduce the flux set, and this in turn couples into
the estimates for component temperatures (through reddening) and bolometric
fluxes (which must account for estimated extinction).
To robustly estimate component radiometric parameters and
uncertainties, we directly evaluated a large grid (roughly $10^5$
cases) spanning the range of viable component temperatures, surface gravities,
and system extinction.   Those ensemble results were then used to seed
Monte Carlo simulations of radiometric parameter a posteriori distributions,
and corresponding angular diameter estimates for the stars via
$\phi^{2}_{\rm rad} = 4 f_{\rm Bol} / \sigma T_{\rm eff}^4$, where $\sigma$
is the Stefan-Boltzmann constant. Figure~\ref{fig:sedcorner}
illustrates the posterior distributions for the visual extinction
and individual component bolometric fluxes, which are the properties
showing the strongest correlations.

We note here that our radiometric diameter $\phi_{\rm rad}$ is not necessarily the
same as the wavelength-dependent angular diameters derived
interferometrically (even after accounting for limb darkening), such as from MIRC or PTI, although both
tell us something about the size of the star. It has been pointed
out previously \citep[see, e.g.,][]{Mihalas:1990, Baschek:1991, Scholz:1997} that
the ``radius" of a star is not a well defined quantity, as it depends
on how it is measured, particularly for giants.
In this paper, we follow the practice of other authors
\citep[e.g.,][]{Hofmann:1998a, Hofmann:1998b, Wittkowski:2004,
Wittkowski:2006a, Wittkowski:2006b} and
interpret our radiometric angular diameter to be a
measure of the size of a star at a Rosseland optical depth of unity.
This definition of the radius is commonly used in atmospheric modeling,
and in formulating the boundary conditions for interior models.

\begin{figure}
\includegraphics[scale=0.34,angle=270]{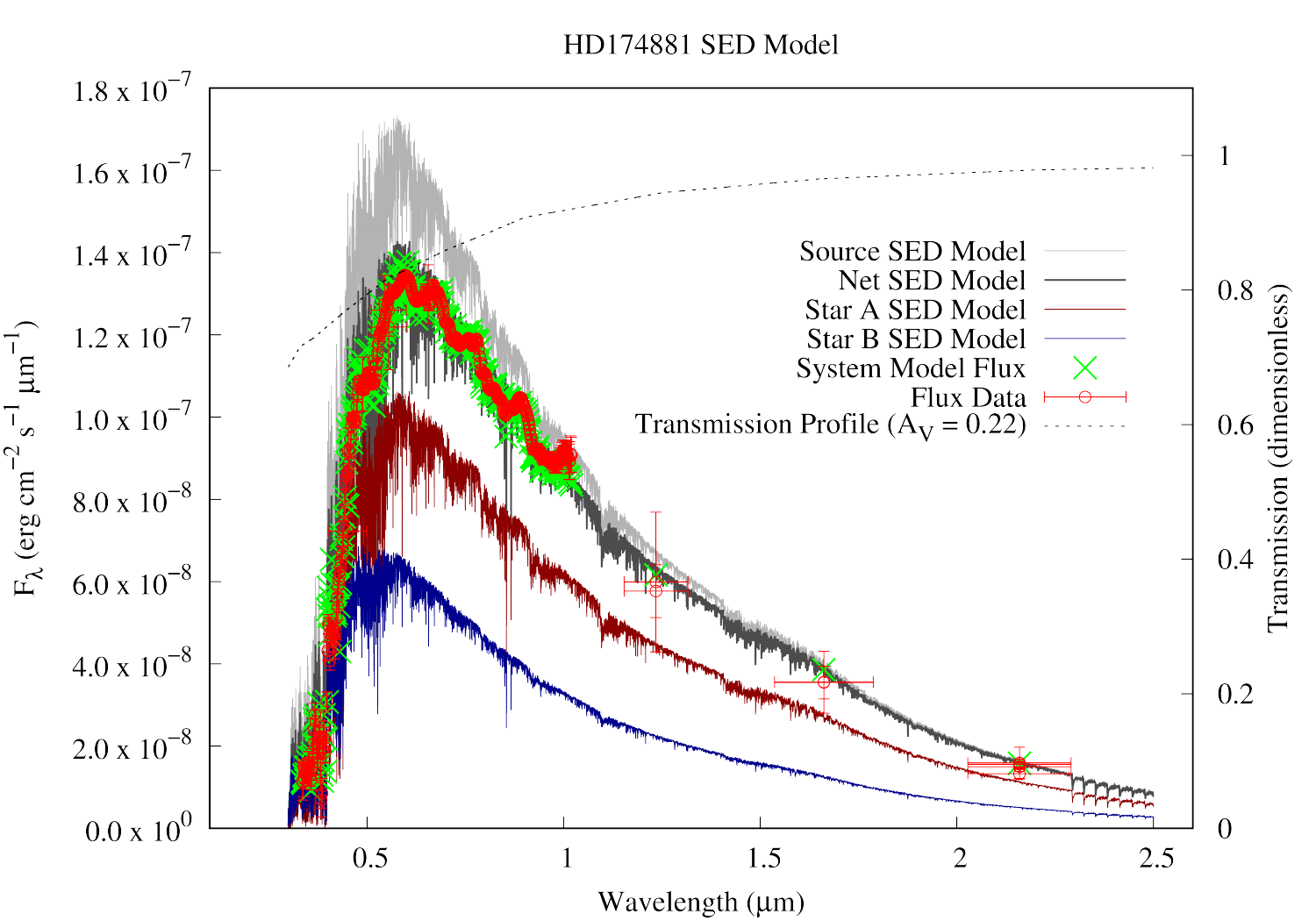}
\figcaption{Spectral energy distribution of \hstar.\label{fig:sed}}
\end{figure}

Independent SED-estimated component parameter values for \hstar\ 
were found to be in good
agreement with the spectroscopic analysis from Section~\ref{sec:spectroscopy}.
Therefore, our preferred a posteriori distributions incorporated Gaussian priors for
the component temperatures ($T_{\rm eff} = 4620 \pm 100$~K and $4880 \pm 150$~K for
stars A and B, respectively), as derived in that section.
Table~\ref{tab:sed} summarizes the results, in support of subsequent
steps in the analysis of the system. The flux measurements along
with our model are presented graphically in Figure~\ref{fig:sed}.

\setlength{\tabcolsep}{10pt}
\begin{deluxetable*}{lccc}
\tablewidth{0pc}
\tablecaption{Radiometric Results for \hstar \label{tab:sed}}
\tablehead{
\colhead{Parameter} &
\colhead{Spectroscopy} &
\colhead{SED, no $T_{\rm eff}$ priors} &
\colhead{SED, with $T_{\rm eff}$ priors}
}
\startdata
\multicolumn{4}{c}{Star A} \\ [0.5ex]
\hline \\ [-1.5ex]
$T_{\rm eff}$ (K)                                 &  $4620 \pm 150$  &  $4631^{+138}_{-149}$    & $4630^{+93}_{-85}$     \\ [0.5ex]
$\phi_{\rm rad}$ (mas)                             &  \nodata         &  $0.78^{+0.04}_{-0.04}$  & $0.78^{+0.03}_{-0.03}$ \\ [1ex]
$f_{\rm bol}$ ($10^{-7}$ erg s$^{-1}$ cm$^{-2}$)  &  \nodata         &  $0.94^{+0.06}_{-0.07}$  & $0.93^{+0.05}_{-0.05}$ \\ [1ex]
$A(V)$ (mag)                                      &  \nodata         &  $0.22^{+0.07}_{-0.07}$  & $0.22^{+0.06}_{-0.06}$ \\ [1ex]
\hline \\ [-1.5ex]
\multicolumn{4}{c}{Star B} \\ [0.5ex]
\hline \\ [-1.5ex]
$T_{\rm eff}$ (K)                                 &  $4880 \pm 150$  &  $4963^{+190}_{-197}$    & $4908^{+133}_{-132}$   \\ [0.5ex]
$\phi_{\rm rad}$ (mas)                             &  \nodata         &  $0.51^{+0.04}_{-0.05}$  & $0.52^{+0.03}_{-0.04}$ \\ [1ex]
$f_{\rm bol}$ ($10^{-7}$ erg s$^{-1}$ cm$^{-2}$)  &  \nodata         &  $0.51^{+0.08}_{-0.06}$  & $0.51^{+0.06}_{-0.06}$ \\ [1ex]
$A(V)$ (mag)                                      &  \nodata         &  $0.22^{+0.07}_{-0.07}$  & $0.22^{+0.06}_{-0.06}$ 

\enddata
\tablecomments{Our adopted SED solution is the one that includes the temperature priors.}
\end{deluxetable*}
\setlength{\tabcolsep}
{6pt}

\begin{figure}
\includegraphics[scale=0.36,angle=270,trim=0in 0.4in 0in 0in]{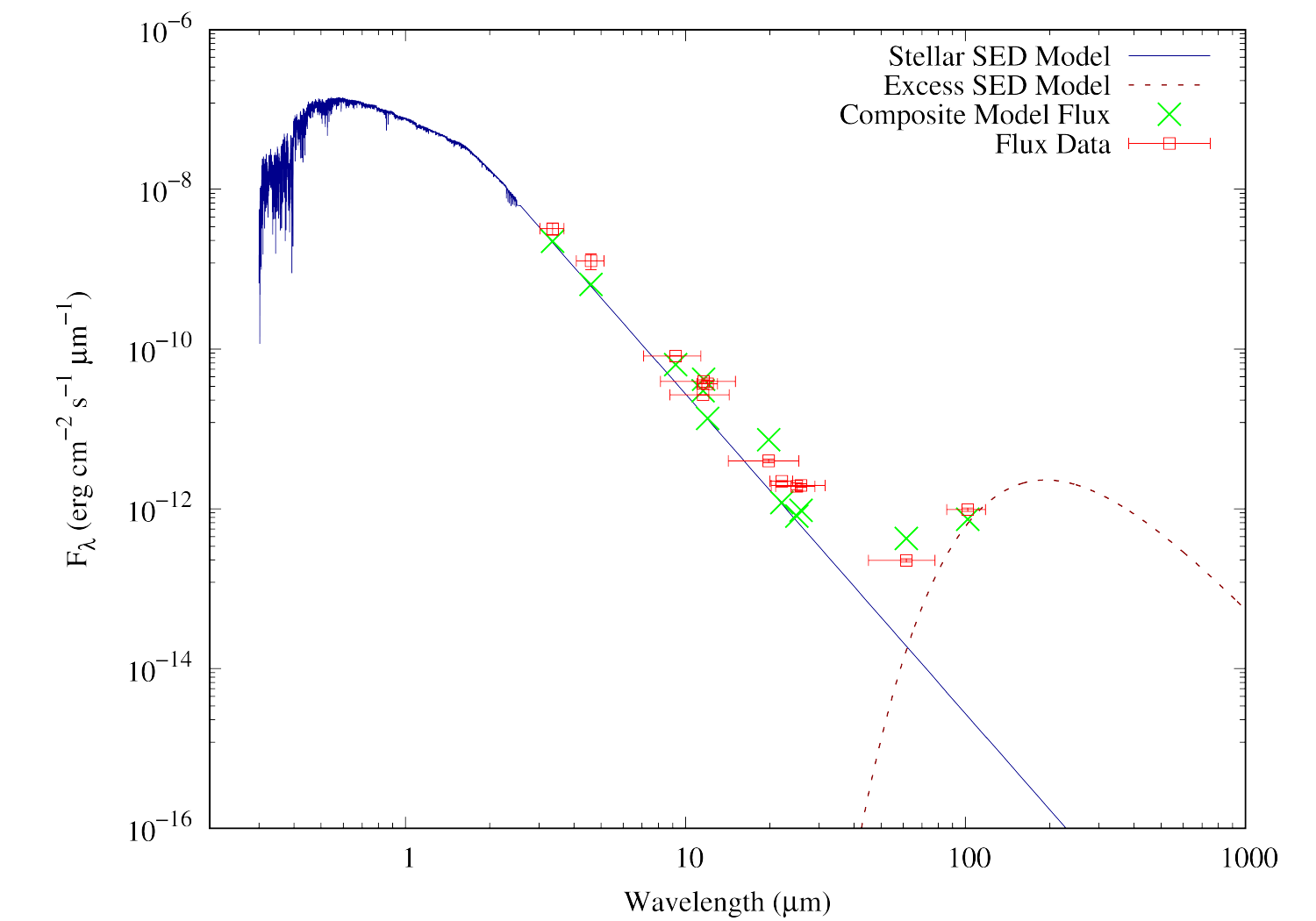}
\figcaption{Spectral energy distribution for \hstar\ extended to
the infrared via a Rayleigh-Jeans approximation, showing an apparent
flux excess longward of about 5\thinspace$\mu$m.
A blackbody model corresponding to a temperature of 15~K provides
a reasonable representation of that excess.
\label{fig:sedir}}
\end{figure}

During our SED analysis, it became apparent that archival flux measurements for \hstar\ beyond around 5\thinspace$\mu$m exhibited excess flux relative to photospheric expectations.
The situation is depicted in Figure~\ref{fig:sedir}, which shows long-wavelength flux measurements plotted against the (sum of Rayleigh-Jeans extensions for) stellar component SEDs.  Modest IR excess flux is apparent over most of the range from 5--50\thinspace$\mu$m.  Apparent excesses in the IRAS bands at 60 and 100\thinspace$\mu$m are much more dramatic.  A simplistic blackbody model fit against these apparent excesses would suggest significant amounts of cool (15--20~K) dust in the \hstar\ system, and this would also seem consistent with the levels of apparent extinction in the radiometric modeling (Table~\ref{tab:sed}). Further investigation of this IR excess will be the topic of ongoing study.

\section{Orbital Solution}
\label{sec:orbit}

The PTI observations and radial-velocity measurements from our own
observations, as well as those of \cite{DeMedeiros:1999}, were
analyzed together to derive the astrometric and spectroscopic orbital
parameters of \hstar\ simultaneously.  The usual elements are the
orbital period ($P$), a reference time of periastron passage ($T_{\rm
 peri}$), the eccentricity ($e$) and argument of periastron for
star~A ($\omega_{\rm A}$), the velocity semiamplitudes ($K_{\rm A}$, $K_{\rm B}$), the
center-of-mass velocity ($\gamma$), the angular semimajor axis
($a^{\prime\prime}$), the inclination angle ($i$), and the position
angle of the ascending node ($\Omega$). The $K$-band flux ratio
$(F_{\rm B}/F_{\rm A})_K$, which is constrained by the PTI observations, is an
additional parameter in this case. The angular diameters of the
components also need to be specified (see below).
For convenience,
the eccentricity and $\omega_{\rm A}$ were recast for our analysis as
$\sqrt{e}\cos\omega_{\rm A}$ and $\sqrt{e}\sin\omega_{\rm A}$ \citep[see,
  e.g.,][]{Anderson:2011, Eastman:2013}, and the inclination angle as
$\cos i$.  We also allowed for a possible systematic shift
($\Delta_{\rm RV}$) between the \cite{DeMedeiros:1999} velocities and
our own.

The analysis was carried out in a Markov Chain Monte Carlo framework
using the {\sc
  emcee}\footnote{\url{https://emcee.readthedocs.io/en/stable/index.html}}
package of \cite{Foreman-Mackey:2013}. We applied uniform priors over
suitable ranges for all of the above adjustable parameters. We
verified convergence by visual inspection of the chains, and also
required a Gelman-Rubin statistic of 1.05 or smaller
\citep{Gelman:1992}.

To guard against internal observational errors that may be either too
small or too large, we included additional free parameters
representing multiplicative scaling factors $f$ for all uncertainties,
separately for the radial velocities of star~A and star~B from the CfA
and from \cite{DeMedeiros:1999}, as well as for the squared visibilities
from the PTI. These scale factors were solved for simultaneously and
self-consistently with the other free parameters
\citep[see][]{Gregory:2005}, using log-uniform priors.

Estimates of the angular size of star~A, from MIRC and
Section~\ref{sec:sed}, indicated it may be resolved by
the PTI. We therefore added its (uniform disk) angular diameter
$\phi_{\rm UD,A}$ as a freely adjustable parameter. The other
component, on the other hand, is too small to be resolved.
Nevertheless, rather than holding the value of $\phi_{\rm UD,B}$
fixed, we allowed it to vary within wide ranges,
subject to priors based on the results from MIRC and Section~\ref{sec:sed}.
The motivation for this is to allow any uncertainty in $\phi_{\rm UD,B}$
to propagate through the analysis to all other parameters.
The MIRC diameter was measured in the $H$ band, rather than $K$,
and estimating what it would translate to in $K$ would require detailed
modeling for \hstar\ to account for differences in opacities and other atmospheric
properties at both wavelengths. As we expect the difference to be
smaller than the formal uncertainty, here we have simply chosen to
use the $H$-band value to establish a Gaussian $K$-band prior from MIRC.
And as our wavelength-independent $\phi_{\rm rad}$ value from
Section~\ref{sec:sed} is not directly related to what is measured
by the PTI, in this case we chose to define a loose uniform prior with a
conservative 3$\sigma$ half width. Both of these priors were
applied simultaneously to determine $\phi_{\rm UD,B}$.

An initial solution for the uniform-disk diameter of star~A
produced the value $\phi_{\rm UD,A} = 0.803^{+0.020}_{-0.025}$~mas.
This of the same order as our earlier estimates from MIRC
($0.817 \pm 0.030$~mas) and from the independent radiometric
analysis of Section~\ref{sec:sed} ($0.78 \pm 0.03$~mas).
Having verified that there are no serious disagreements, for our
final MCMC solution we incorporated the information from latter
two measurements by applying priors on $\phi_{\rm UD,A}$ in the
same way as done above for star~B. As the main goal of this analysis
was to derive accurate orbital parameters for \hstar, $\phi_{\rm UD,A}$
and $\phi_{\rm UD,B}$ are regarded here merely as nuisance parameters.
The different estimates of the angular diameters for stars~A and B
are summarized in Table~\ref{tab:diam}.

\setlength{\tabcolsep}{8pt}
\begin{deluxetable}{lcc}
\tablewidth{0pc}
\tablecaption{Apparent Angular Diameter estimates for \hstar \label{tab:diam}}
\tablehead{
\colhead{Source} &
\colhead{$\phi_{\rm A}$} &
\colhead{$\phi_{\rm B}$}
\\
\colhead{} &
\colhead{(mas)} &
\colhead{(mas)}
}
\startdata
MIRC ($\phi_{\rm UD})$                    & $0.817 \pm 0.030$       &  $0.50 \pm 0.05$           \\
SED Analysis ($\phi_{\rm rad})$           & $0.78^{+0.03}_{-0.03}$  &  $0.52^{+0.03}_{-0.04}$           \\ [+1ex]
Initial PTI Analysis\tablenotemark{a} ($\phi_{\rm UD}$)    & $0.803^{+0.020}_{-0.025}$  &  $0.534^{+0.044}_{-0.052}$ \\ [1ex]
Final PTI Analysis\tablenotemark{b} ($\phi_{\rm UD}$)      & $0.807^{+0.018}_{-0.018}$  &  $0.521^{+0.047}_{-0.044}$ 
\enddata
\tablenotetext{a}{This MCMC analysis imposed simultaneous priors on
$\phi_{\rm B}$ based on results from MIRC and the SED (see the text),
but left $\phi_{\rm A}$ completely free. It was meant to
verify that the constraint on the diameter of star~A from the PTI alone is
consistent with the estimates from the two methods above. The exercise
proved that to be the case.}
\tablenotetext{b}{These estimates used the same priors on $\phi_{\rm B}$
as above, and incorporated the information from MIRC and the SED fit in
the form of additional priors on $\phi_{\rm A}$.}
\end{deluxetable}
\setlength{\tabcolsep}{6pt}

\begin{figure}
\epsscale{1.22}
\includegraphics[width=9.2cm,trim={15 5 10 0},clip]{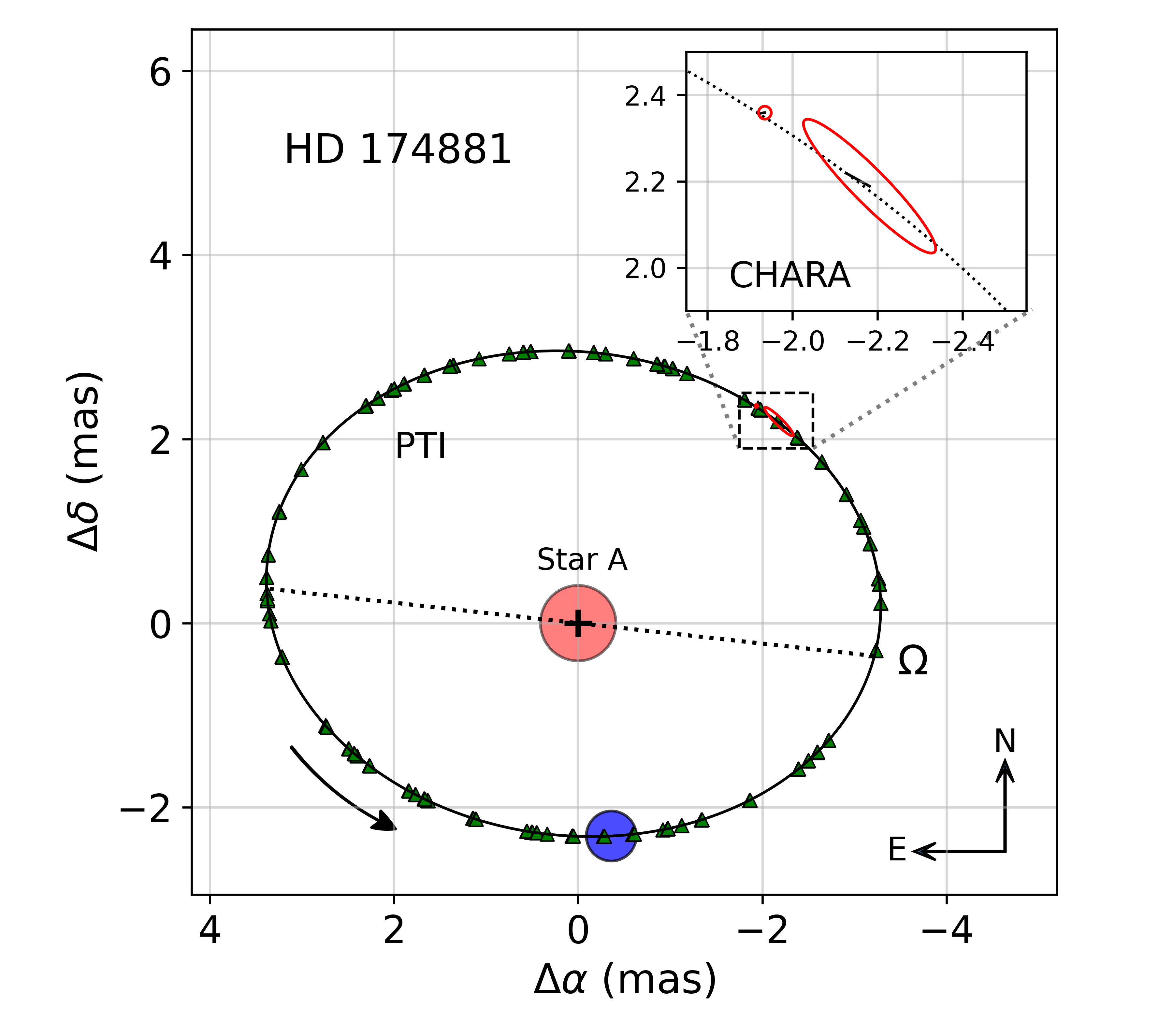}
\figcaption{Astrometric model for \hstar. Neither the PTI visibilities nor
their uncertainties can be plotted on the plane of the sky, but we
represent them here as green triangles at their predicted locations
in the orbit, to illustrate the phase coverage they provide.
The components are drawn with their sizes
to scale relative to the orbit, and star~B (the smaller and more
massive component, in blue) is rendered at periastron.
The dotted line marks the line of nodes, and the
ascending node is indicated with the ``$\Omega$" symbol.
In accordance with the classical convention, $\Omega$ is the node at
which star~B is receding from the observer relative to $\gamma$.
The enlargement in the inset shows the two archival CHARA observations,
with their corresponding error ellipses. The short line segments
connecting the measured and predicted CHARA positions represent the
residuals from the model. \label{fig:pti}}
\end{figure}

\begin{figure}
\epsscale{1.15}
\plotone{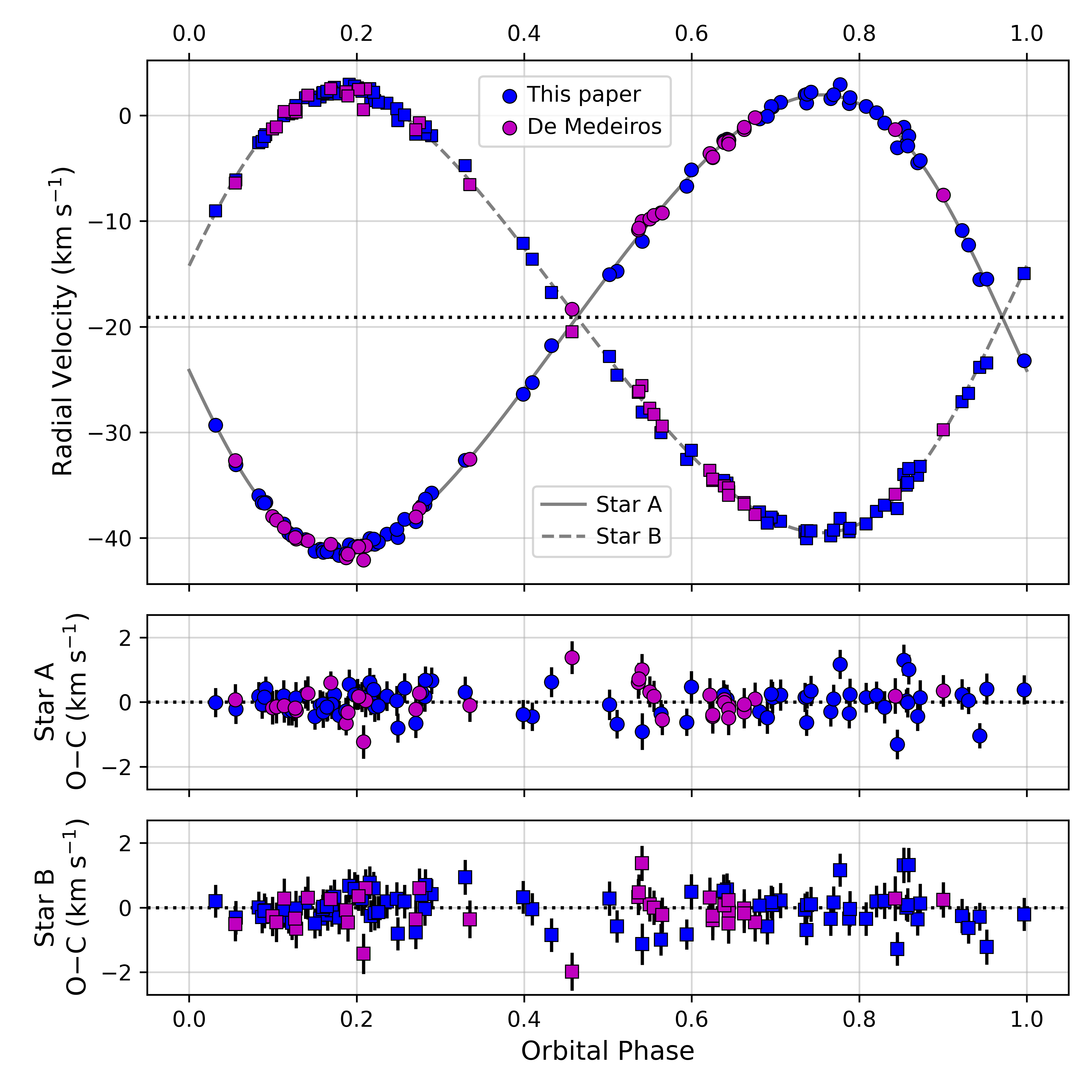}
\figcaption{Spectroscopic orbit for \hstar, together with
our RV measurements and those of \cite{DeMedeiros:1999}.
The dotted line in the top panel marks the center-of-mass
velocity of the system. Residuals are shown at the bottom.
  \label{fig:rvs}}
\end{figure}

\begin{figure}
\includegraphics[angle=270,scale=0.34]{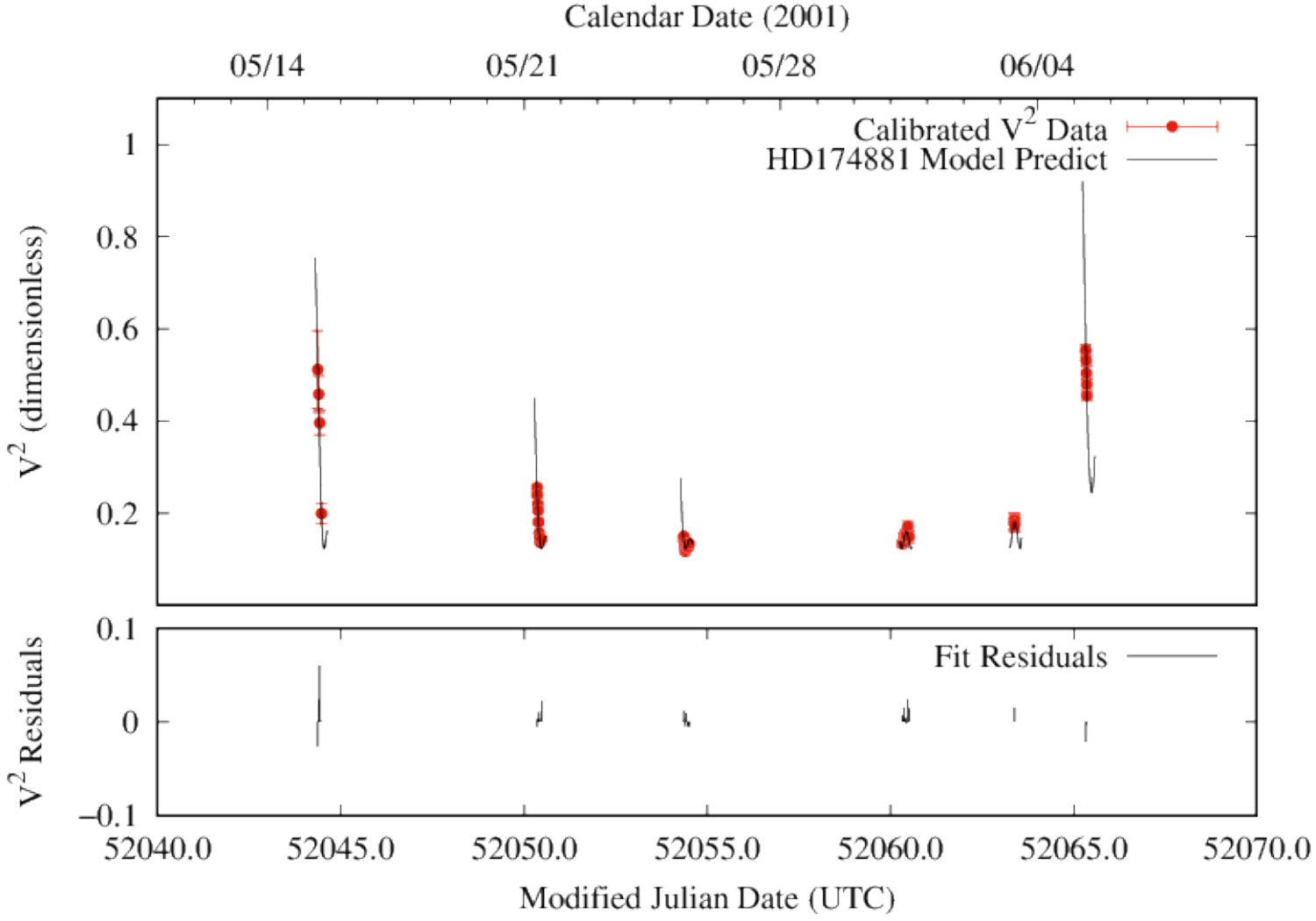}
\vskip 10pt
\includegraphics[angle=270,scale=0.34]{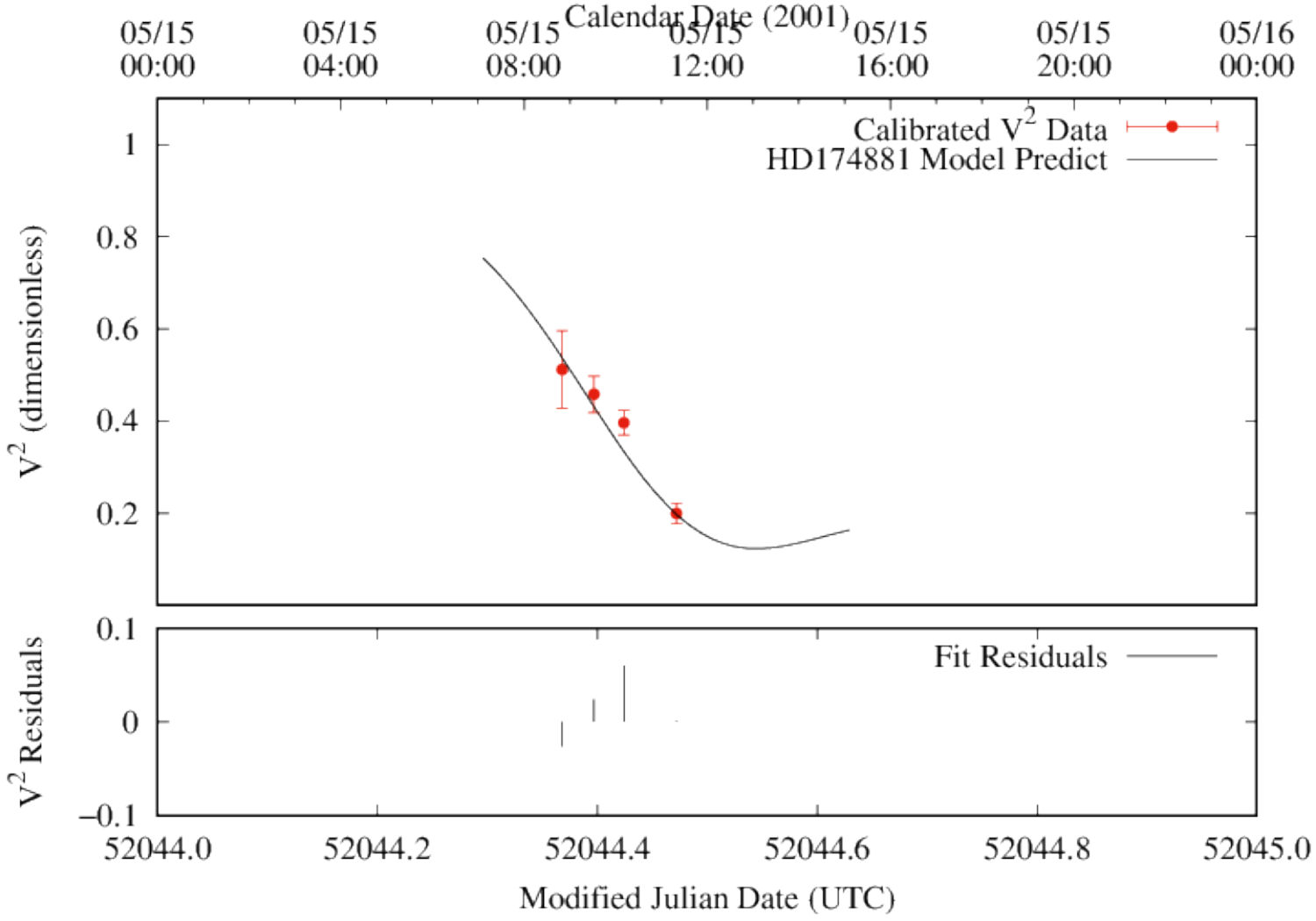}
    \figcaption{Visualization of the fit of our orbit model to
    the calibrated PTI visibilities for \hstar. The top panel shows
    the measurements and model for six nights over a one month stretch in 2001,
    and the bottom panel presents an enlargement of the first of those
    nights (2001 May 15). The visibility residuals are shown in
    both panels.\label{fig:PTIfit}}
\end{figure}

The complete results of our orbital analysis are presented in Table~\ref{tab:mcmc}. 
Our astrometric orbit model is
shown in Figure~\ref{fig:pti}, in which the PTI $V^2$ measurements,
which cannot be represented in this plot, are shown as triangles at
their predicted locations. The inset displays the two archival CHARA observations,
which were not included in the fit but match the predicted relative
positions well within their uncertainties. The radial velocities are
shown with the spectroscopic orbit in Figure~\ref{fig:rvs}.
An illustration of the fit to the PTI visibilities is shown in
Figure~\ref{fig:PTIfit}.

\setlength{\tabcolsep}{4pt}
\begin{deluxetable}{lcc}
\tablewidth{0pc}
\tablecaption{Results of our Orbital Analysis for \hstar \label{tab:mcmc}}
\tablehead{
\colhead{~~~~~~~~~~~Parameter~~~~~~~~~~~} &
\colhead{Value} &
\colhead{Prior}
}
\startdata
 $P$ (day)                          & $215.1166^{+0.0092}_{-0.0072}$            & [100, 300]      \\ [1ex]
 $T_{\rm peri}$ (HJD$-$2,400,000)   & $51843.62^{+0.20}_{-0.17}$                & [51800, 51900]  \\ [1ex]
 $\sqrt{e}\cos\omega_{\rm A}$             & $-0.0739^{+0.0019}_{-0.0021}$       & [$-1$, 1]       \\ [1ex]
 $\sqrt{e}\sin\omega_{\rm A}$             & $+0.3409^{+0.0013}_{-0.0010}$       & [$-1$, 1]       \\ [1ex]
 $a^{\prime\prime}$ (mas)           & $3.3684^{+0.0060}_{-0.0071}$              & [1, 10]         \\ [1ex]
 $\cos i$                           & $0.7801^{+0.0019}_{-0.0018}$              & [$-1$, 1]       \\ [1ex]
 $\Omega$ (deg)                     & $263.65^{+0.15}_{-0.15}$                  & [0, 360]        \\ [1ex]
 $\gamma$ (\kms)                    & $-19.101^{+0.034}_{-0.043}$               & [$-30$, 0]      \\ [1ex]
 $K_{\rm A}$ (\kms)                       & $21.594^{+0.053}_{-0.053}$          & [10, 50]        \\ [1ex]
 $K_{\rm B}$ (\kms)                       & $20.924^{+0.062}_{-0.057}$          & [10, 50]        \\ [1ex]
 $\Delta_{\rm RV}$ (\kms)           & $+0.193^{+0.082}_{-0.064}$                & [$-5$, 5]       \\ [1ex]
 $(F_{\rm B}/F_{\rm A})_K$                      & $0.4594^{+0.0022}_{-0.0026}$  & [0.1, 3.0]      \\ [1ex]
 $\phi_{\rm UD,A}$ (mas)                     & $0.807^{+0.018}_{-0.018}$        & $G \times U$    \\ [1ex]
 $\phi_{\rm UD,B}$ (mas)                     & $0.521^{+0.047}_{-0.044}$        & $G \times U$    \\ [1ex]
 $f_{\rm CfA,A}$, $\sigma_{\rm A}$ (\kms) & $0.984^{+0.091}_{-0.066}$, 0.50     & [$-5$, 5]       \\ [1ex]
 $f_{\rm CfA,B}$, $\sigma_{\rm B}$ (\kms) & $0.988^{+0.100}_{-0.061}$, 0.45     & [$-5$, 5]       \\ [1ex]
 $f_{\rm DM,A}$, $\sigma_{\rm A}$ (\kms)  & $1.10^{+0.16}_{-0.11}$, 0.56        & [$-5$, 5]       \\ [1ex]
 $f_{\rm DM,B}$, $\sigma_{\rm B}$ (\kms)  & $1.17^{+0.17}_{-0.11}$, 0.47        & [$-5$, 5]       \\ [1ex]
 $f_{\rm PTI}$, $\sigma_{V^2}$      & $1.106^{+0.042}_{-0.031}$, 0.016          & [$-5$, 5]       \\ [1ex]
\hline \\ [-1.5ex]
\multicolumn{3}{c}{Derived quantities} \\ [1ex]
\hline \\ [-1.5ex]
 $e$                                & $0.12162^{+0.00063}_{-0.00064}$           & \nodata        \\ [1ex]
 $\omega_{\rm A}$ (deg)                   & $101.98^{+0.36}_{-0.34}$            & \nodata        \\ [1ex]
 $i$ (deg)                          & $38.73^{+0.17}_{-0.17}$                   & \nodata        \\ [1ex]
 Total mass ($M_{\sun}$)            & $6.838^{+0.093}_{-0.079}$                 & \nodata        \\ [1ex]
 $M_{\rm A}$ ($M_{\sun}$)                 & $3.367^{+0.045}_{-0.041}$           & \nodata        \\ [1ex]
 $M_{\rm B}$ ($M_{\sun}$)                 & $3.476^{+0.043}_{-0.043}$           & \nodata        \\ [1ex]
 $q \equiv M_{\rm B}/M_{\rm A}$     & $1.0318^{+0.0040}_{-0.0037}$              & \nodata        \\ [1ex]
 $a$ (au)                   & $1.3336^{+0.0060}_{-0.0051}$                      & \nodata        \\ [1ex]
 $\pi_{\rm orb}$ (mas)              & $2.525^{+0.014}_{-0.015}$                 & \nodata        \\ [1ex]
 Distance (pc)                      & $396.0^{+2.4}_{-2.2}$                     & \nodata        \\ [1ex]
 $R_{\rm A}$ ($R_{\sun}$)                 & $34.0^{+1.3}_{-1.3}$                & \nodata        \\ [1ex]
 $R_{\rm B}$ ($R_{\sun}$)                 & $22.7^{+1.8}_{-1.8}$                & \nodata        \\ [1ex]
 $\log g_{\rm A}$ (cgs)                   & $1.903^{+0.033}_{-0.033}$           & \nodata        \\ [1ex]
 $\log g_{\rm B}$ (cgs)                   & $2.262^{+0.075}_{-0.059}$           & \nodata        
\enddata

\tablecomments{The values listed correspond to the mode of the
  posterior distributions, and the uncertainties are the 68.3\%
  credible intervals. $f_{\rm CfA,A}$ and $f_{\rm CfA,B}$ are the
  scale factors for the internal errors of the CfA RV velocities of
  the two components. A similar notation is used for the RVs of
  \cite{DeMedeiros:1999}, and the PTI. Values following these scale
  factors on the same line are the weighted rms residuals, after
  application of the scale factors. Priors in square brackets are
  uniform over the ranges specified, except for those of the error
  scaling factors $f$, which are log-uniform.
  For $\phi_{\rm UD,A}$ and $\phi_{\rm UD,B}$, the $G \times U$
  notation indicates the product of Gaussian and Uniform priors
  as described in the text.
  All derived quantities in the bottom section of the table were
  computed directly from the Markov chains of the fitted parameters
  involved. The absolute radii depend on the radiometric angular
  diameters $\phi_{\rm rad}$ from Section~\ref{sec:sed}, and
  our distance. With the masses, we then computed $\log g$.}

\end{deluxetable}
\setlength{\tabcolsep}{6pt}

We note that the position angle of the ascending node ($\Omega$), as
determined from the combination of PTI measurements and radial
velocities, still suffers from a 180\arcdeg\ ambiguity due to the
fact that the interferometric squared visibilities are invariant
under a point-symmetric inversion around the binary origin. The
MIRC observations of \hstar\ break that degeneracy.

The Gaia mission has reported a spectroscopic orbit for \hstar\
(source ID 2040514502502017536)
in its most recent data release \citep[DR3;][]{GaiaDR3:2023}, in which the velocities
of both components were measured. The elements are reproduced
in Table~\ref{tab:gaia}, for easier comparison with the results
of this paper. While the Gaia orbit is largely correct, several of
the elements show significant deviations from our more precise values.

Our inferred distance for \hstar, $396.0^{+2.4}_{-2.2}$~pc, is in excellent
agreement with the value derived from the Gaia DR3 parallax, after
adjusting it for the zeropoint offset reported by \cite{Lindegren:2021}.
The Gaia value is $397.4 \pm 5.0$~pc.

\setlength{\tabcolsep}{10pt}
\begin{deluxetable}{lc}
\tablewidth{0pc}
\tablecaption{Spectroscopic Orbital Solution for \hstar\ from Gaia DR3  \label{tab:gaia}}
\tablehead{
\colhead{Parameter} &
\colhead{Value}
}
\startdata
$P$ (days)                       & $215.654 \pm 0.063$\phn\phn \\
$T_{\rm peri}$ (HJD$-$2,400,000)\tablenotemark{a} & $57421.82 \pm 0.97$\phm{2222} \\
$e$                              & $0.1220 \pm 0.0025$ \\
$\omega_{\rm A}$ (deg)           & $75.4 \pm 1.6$\phn      \\
$\gamma$ (\kms)                  & $-19.420 \pm 0.035$\phn\phs \\
$K_{\rm A}$ (\kms)               & $21.531 \pm 0.075$\phn  \\
$K_{\rm B}$ (\kms)               & $19.280 \pm 0.073$\phn  
\enddata
\tablenotetext{a}{This time of periastron passage is shifted forward
by 26 orbital cycles from our value in Table~\ref{tab:mcmc}. Adjusting it
backwards using our more precise period gives 51828.76.}
\end{deluxetable}
\setlength{\tabcolsep}{6pt}

\section{Discussion}
\label{sec:discussion}

The main properties of the \hstar\ stars are collected in
Table~\ref{tab:properties}. The fainter component (star~B) is
the more massive one, and is therefore more evolved. 
It would typically be referred to as the ``primary" in the system.
It is smaller and hotter than its companion.

\setlength{\tabcolsep}{10pt}
\begin{deluxetable}{lcc}
\tablewidth{0pc}
\tablecaption{Summary of the Physical Properties of \hstar \label{tab:properties}}
\tablehead{
\colhead{Parameter} &
\colhead{Star A} &
\colhead{Star B}
}
\startdata
$M$ ($M_{\sun}$)        & $3.367^{+0.045}_{-0.041}$  & $3.476^{+0.043}_{-0.043}$ \\ [1ex]
$R$ ($R_{\sun}$)        & $34.0^{+1.3}_{-1.3}$       & $22.7^{+1.8}_{-1.8}$      \\ [1ex]
$\log g$ (cgs)          & $1.903^{+0.033}_{-0.033}$  & $2.262^{+0.075}_{-0.059}$ \\ [1ex]
$T_{\rm eff}$ (K)       & $4620 \pm 100$\phn         & $4880 \pm 150$\phn        \\
$\log L/L_{\sun}$       & $2.659 \pm 0.024$          & $2.398 \pm 0.052$         \\ [0.5ex]
Distance (pc)           & \multicolumn{2}{c}{$396.0^{+2.4}_{-2.2}$}              \\ [1ex]
$A(V)$ (mag)            & \multicolumn{2}{c}{$0.22^{+0.06}_{-0.06}$}             \\ [0.5ex]
$M_V$ (mag)             & $-1.462 \pm 0.065$\phs     & $-1.042 \pm 0.066$\phs    \\
$M_H$ (mag)             & $-3.87 \pm 0.23$\phs       & $-3.02 \pm 0.23$\phs      \\
$M_{K_{\rm S}}$ (mag)   & $-3.984 \pm 0.063$\phs     & $-3.138 \pm 0.064$\phs    
\enddata

\tablecomments{The sources of the above properties are as follows: $M$, $R$, $\log g$, and the distance are taken from Table~\ref{tab:mcmc}.
The temperatures are spectroscopic (Section~\ref{sec:spectroscopy}). The luminosities rely on the bolometric fluxes from
the radiometric analysis, and the distance. Extinction also comes from the radiometric analysis. The absolute magnitudes
depend on the system magnitudes in each bandpass, extinction, the measured flux ratios, and the distance (see the text).}
\end{deluxetable}
\setlength{\tabcolsep}{6pt}

Here we compare these properties against two sets of recent
stellar evolution models, under the assumption that the components
are coeval. To further constrain the models, we have added to the
wavelength-independent dynamical properties and the bolometric luminosities
the (extinction-corrected) absolute magnitudes of
the components in the $V$, $H$, and $K_{\rm S}$ bandpasses. They depend on
the combined-light brightness, the in-band flux ratios, and our distance estimate.
The flux ratio in the $V$ band was obtained by applying a small
correction to our spectroscopic value reported in Section~\ref{sec:spectroscopy},
with the aid of PHOENIX model spectra from \cite{Husser:2013} appropriate for
the two stars. We obtained ($F_{\rm B}/F_{\rm A})_V = 0.679 \pm 0.020$.
The values in the near infrared have been reported earlier, and are
($F_{\rm B}/F_{\rm A})_H = 0.458 \pm 0.010$ and
($F_{\rm B}/F_{\rm A})_K = 0.459 \pm 0.010$, with slightly more
conservative uncertainties adopted here than the nominal ones.
The apparent magnitude of the system in the visual band was
taken to be $V = 6.18 \pm 0.02$ \citep{Mermilliod:1987}.
Due to its near-infrared brightness, \hstar\ is saturated in the 2MASS $H$ and
$K_{\rm S}$ bands. For $H$, we had little choice but to adopt
the 2MASS value as published ($H = 3.74 \pm  0.23$), with its correspondingly
large uncertainty.
For the $K_{\rm S}$ band, we were able to gather new measurements with the generous
help of our colleague Cullen Blake, on the nights of 2006 November~2
and November~8. These observations were made on the 1.3m PAIRITEL telescope
\citep{Blake:2008}, located at the Fred L.\ Whipple Observatory, which was
equipped with the same near-infrared camera that was originally used for
the southern portion of the 2MASS Survey. The average for the two nights
is $K_{\rm S} = 3.62 \pm 0.06$, which is consistent with, but more precise
than, the original 2MASS value of $3.58 \pm 0.26$.

\begin{figure*}
\epsscale{1.17}
\plotone{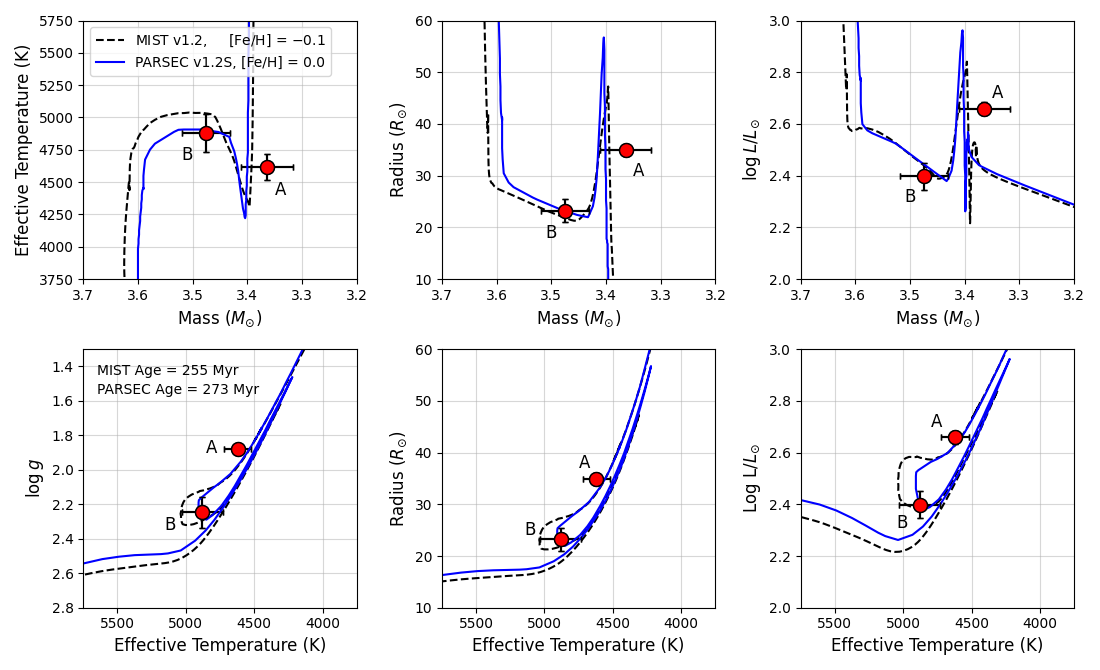}
\figcaption{Comparison of the observed properties of \hstar\ against
model isochrones from the MIST series \citep{Choi:2016} and
the PARSEC series \citep{Chen:2014}. Rotation is not considered.
The panels display the physical properties of the
components (labeled) as a function of mass and temperature.
The best compromise between the observations and theoretical predictions
is reached for a metallicity of ${\rm [Fe/H]} = -0.1$ for MIST, and
solar composition for PARSEC. The corresponding ages are 255 and 273~Myr,
respectively.
\label{fig:modelsa}}
\end{figure*}

\begin{figure*}
\epsscale{1.17}
\plotone{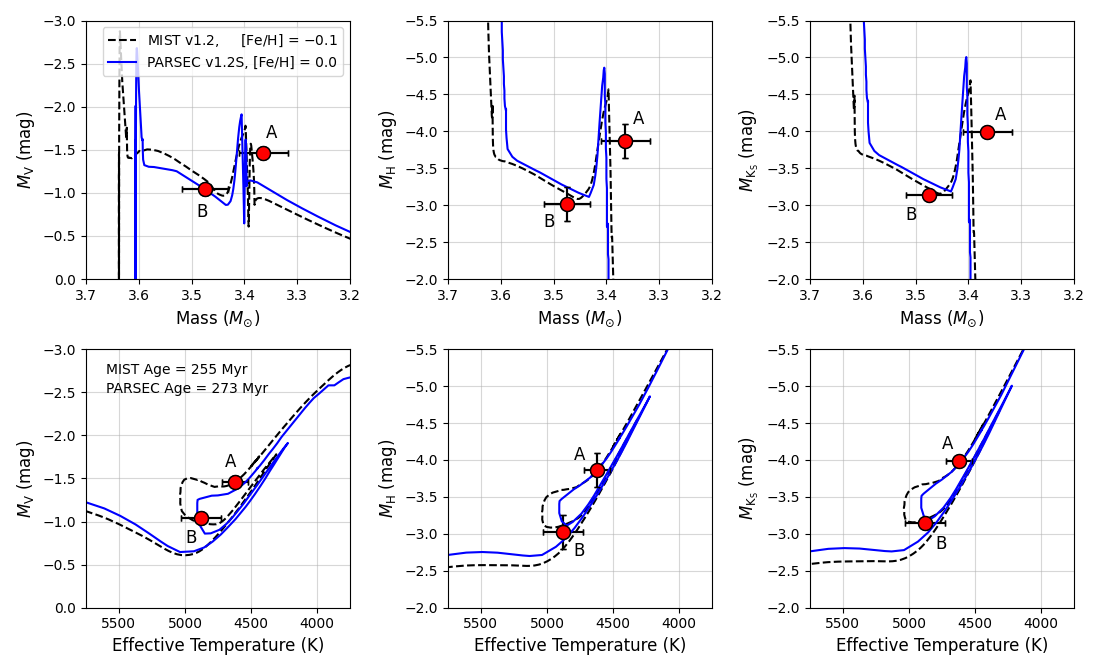}
\figcaption{Similar to Figure~\ref{fig:modelsa}, for the absolute
magnitudes of \hstar\ in $V$, $H$, and $K_{\rm S}$, as a function
of $M$ and $T_{\rm eff}$.
\label{fig:modelsb}}
\end{figure*}

Figure~\ref{fig:modelsa} presents the comparison of the \hstar\ properties
with isochrones from the MIST~v1.2 models of \cite{Choi:2016}, 
as well as the PARSEC~v1.2S models of \cite{Chen:2014}. Both sets of
models use plane-parallel geometry for the atmospheres at the $\log g$ values of the A and B
components, although differences compared to spherical geometry should not be important at these surface gravities.
The six panels illustrate the match to the masses, temperatures, radii, surface
gravities, and bolometric luminosities. The wavelength-dependent absolute
magnitudes ($V$, $H$, $K_{\rm S}$) are shown in Figure~\ref{fig:modelsb},
as a function of mass and effective temperature.
For the MIST models, the best compromise overall was found for a model having a slightly
subsolar composition of ${\rm [Fe/H]} = -0.1$. The corresponding age of
the best fit model is 255~Myr.
For PARSEC, a solar composition isochrone produces a better fit than subsolar,
although it is somewhat worse than the best MIST match. The age in this case is 273~Myr.

Neither model is able to reproduce all eight measured quantities
simultaneously for both stars, within their respective uncertainties. 
In particular, both isochrones predict that star~A
(the less massive and therefore less evolved one) should be
somewhat cooler and/or less luminous than we observe, if its location is to
be consistent with its less evolved state.
Indeed, despite these discrepancies, the precision of the observables is such that
they easily show the two stars to be in
different evolutionary stages. While the more massive
component (star~B) is clearly located in the helium-burning clump, the location of
the other star is either on the first ascent of the giant branch, or on its
way down to the clump. The former position appears more likely, based on the
sum total of the observations (e.g., top panels of Figures~\ref{fig:modelsa}
and \ref{fig:modelsb}).

The progenitors of both components were late B-type stars. Such objects
typically have relatively high initial rotation rates on the zero-age main sequence,
which theory shows can affect their observable properties at later stages of evolution
\citep[see, e.g.,][]{Meynet:1997}.
An additional comparison (not shown) was made against a version of the MIST isochrones that includes the effects of rotation ($\omega/\omega_0 = 0.4$, where $\omega$ here
is the angular rotation rate, and
$\omega_0$ is the value at breakup). The results are rather
similar to the non-rotating case, with the best match to the observations being 
achieved at a marginally older age of 259~Myr.

\section{Conclusions}
\label{sec:conclusions}

Precise, model-independent mass determinations for giant stars
are still relatively uncommon, compared to similar studies for
main-sequence stars. In this paper, we have combined long-baseline
interferometry and high-resolution spectroscopy for the giant system \hstar,
to derive absolute masses with precisions of 1.3\% for both components,
along with a distance (orbital parallax) good to 0.6\%.
We have also determined the absolute radius of the less
massive, larger, and cooler star with an error of just 3.8\%, while
the size of the other star is less well determined (7.9\%). The effective
temperatures of both components have been inferred from spectroscopy.
Additionally, by incorporating flux measurements in a number of
bandpasses, we have derived estimates of the bolometric luminosities
as well as the absolute magnitudes in three different bandpasses
($V$, $H$, $K_{\rm S}$), in ways that
are not completely dependent on the previously determined properties,
thereby adding new information.

In aggregate, these properties provide stringent constraints on
models of stellar evolution for evolved stars. Comparisons against
two sets of current models (MIST v1.2, PARSEC v1.2S) indicate fair
agreement for compositions near solar and ages in the range 255--273~Myr,
although discrepancies remain for some of the measured properties.
The more massive star resides
in the helium-burning clump, while the location in the H--R diagram
of the other, less evolved component is still somewhat ambiguous.
It is either on the first ascent of the giant branch, or on the subsequent
descent toward the clump, the former being favored by the observations.

\acknowledgements 

We are grateful to J.\ Caruso, D.\ W.\ Latham, R.\ P.\ Stefanik, and
J.\ Zajac for their efforts at the telescope to obtain most of the
spectroscopic observations used in this work, and to R.\ J.\ Davis for
maintaining the CfA echelle database.
We also thank Cullen Blake (Univ.\ of Pennsylvania) for
gathering measurements of the $K_{\rm S}$-band magnitude of \hstar\
for this work at our request, and Leo Girardi (INAF) for information
on the PARSEC models. We are grateful as well to the referee for helpful suggestions.
Science operations with the PTI were conducted through the efforts of the
PTI Collaboration, and we acknowledge the invaluable contributions of
our PTI longtime colleagues. 
JDM wishes to thank Ming Zhao, Ettore Pedretti, and Nathalie Thureau for contributions to the MIRC observations, as well as to acknowledge funding from the University of Michigan.
This research has made use of
the SIMBAD and VizieR databases, operated at CDS, Strasbourg, France,
of NASA's Astrophysics Data System Abstract Service, and of data
products from the Two Micron All Sky Survey, which is a joint project
of the University of Massachusetts and the Infrared Processing and
Analysis Center/California Institute of Technology, funded by NASA and
the NSF. 
This work has also made use of data from the European Space Agency (ESA) mission
Gaia (\url{https://www.cosmos.esa.int/gaia}), processed by the Gaia
Data Processing and Analysis Consortium (DPAC,
\url{https://www.cosmos.esa.int/web/gaia/dpac/consortium}). Funding for the DPAC
has been provided by national institutions, in particular the institutions
participating in the Gaia Multilateral Agreement.
The computational resources used for this research include the Smithsonian
High Performance Cluster (SI/HPC), Smithsonian Institution
(\url{https://doi.org/10.25572/SIHPC}).



\section{Appendix: Comparison of our 2007 July 7 MIRC data and the MACIM image model}
\label{sec:appendixA}

The figures in this Appendix compare our MIRC squared visibilities,
closure phases, and triple amplitude measurements from the 2007 July 7 observation with the model for the image of \hstar\ displayed in Figure~\ref{fig:image}. The reduced $\chi^2$ values we obtained from the fit are 0.753 for the visibilities, 0.406 for the closure phases, and 1.571 for the triple amplitudes.

\begin{figure*}
\includegraphics[trim=0.7in 0.9in 0.9in 0.9in,scale=0.98]{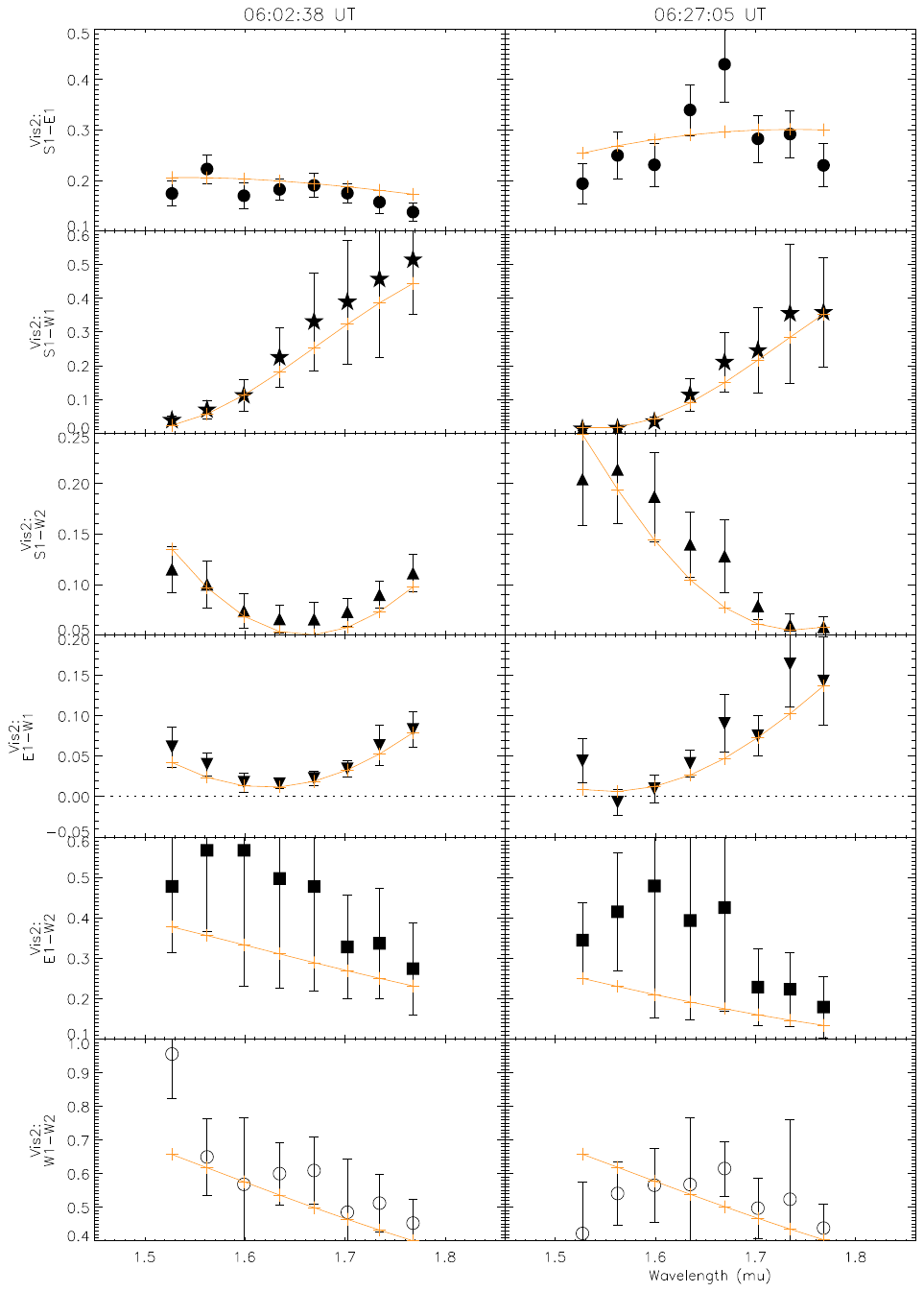}
    \figcaption{Comparison of the measured MIRC squared visibilities from our
    2007 July 7 observation with the model (solid lines) for the MACIM image in
    Figure~\ref{fig:image}.
    \label{fig:MACIMpage6}}
\end{figure*}

\begin{figure*}
\includegraphics[trim=0.7in 0.9in 0.9in 0.9in,scale=0.98]{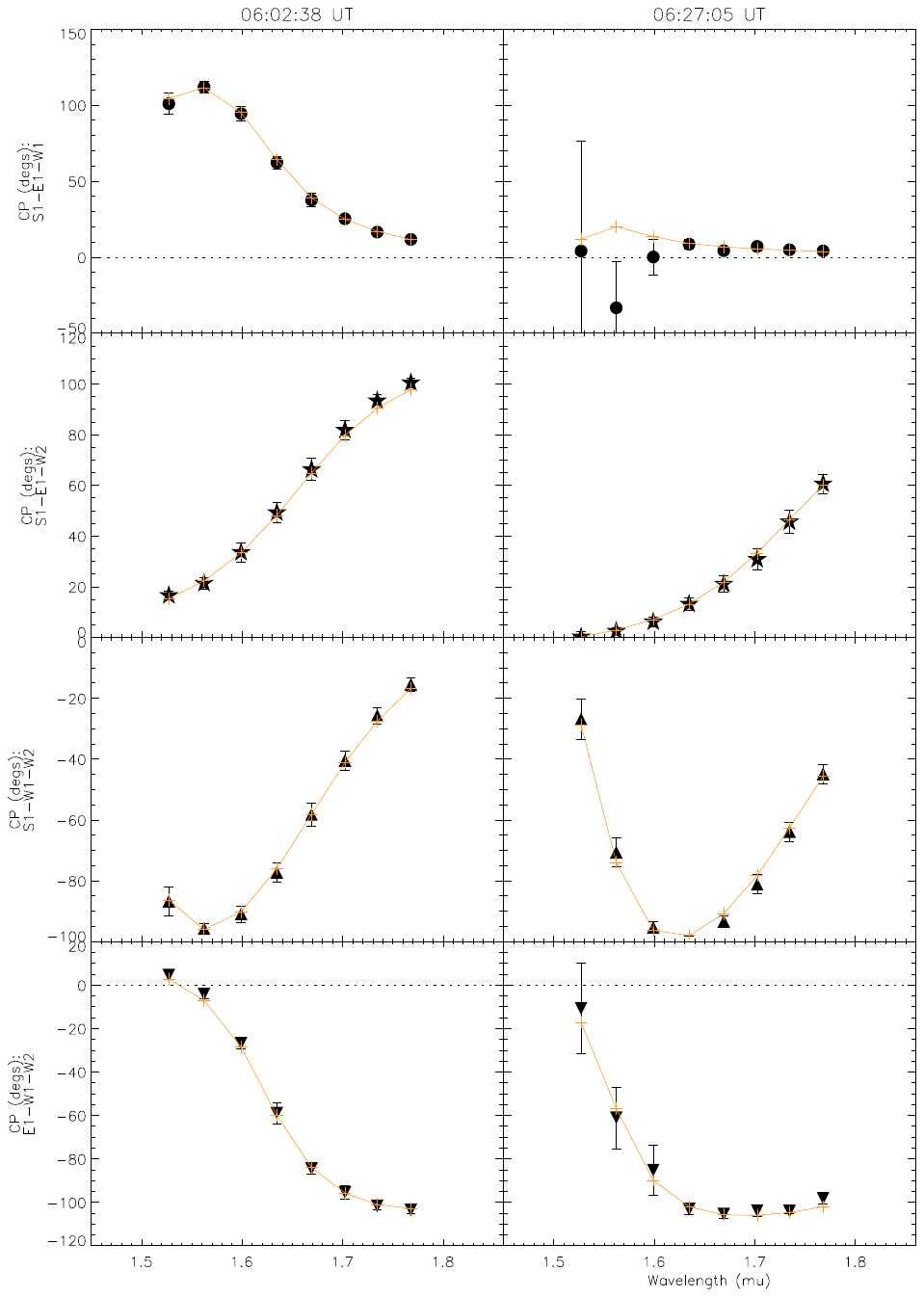}
    \figcaption{Similar to Figure~\ref{fig:MACIMpage6}, for the
    closure phases. \label{fig:MACIMpage7}}
\end{figure*}

\begin{figure*}
\includegraphics[trim=0.7in 0.9in 0.9in 0.9in,scale=0.98]{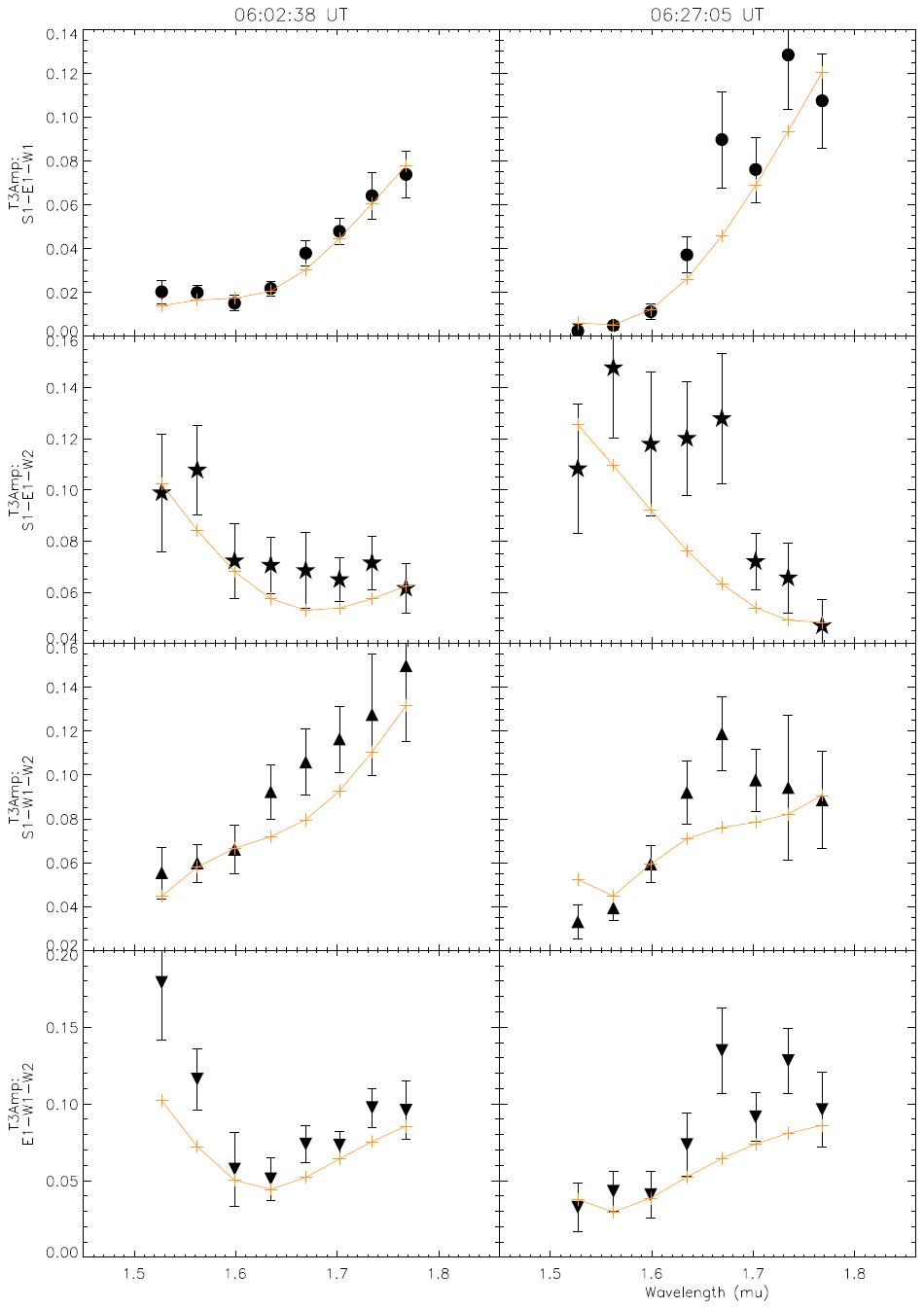}
    \figcaption{Similar to Figure~\ref{fig:MACIMpage6}, for the
    triple product amplitudes. \label{fig:MACIMpage8}}
\end{figure*}




\end{document}